# Automation of Mobile Pick and Place Robotic System for Small food Industry.

Mir Sajjad Hussain Talpur,
School of Information Science and Engineering,
Central South University, Changsha,
Hunan Province, 410083, P.R China.

Murtaza Hussain Shaikh,
Department of Computer and Information Science,
Norwegian University of Science & Technology (NTNU),
Trondheim, 7448 - Norway.

*Abstract*— The use of robotics in food industry is becoming more popular in recent years. The trend seems to continue as long as the robotics technology meets diverse and challenging needs of the food producers. Rapid developments in digital computers and control systems technologies have significant impact in robotics like any other engineering fields. By utilizing new hardware and software tools, design of these complex systems that need strong integration of distinct disciplines is no longer difficult compared to the past. Therefore, the purpose of this paper is to design and implement a microcontroller based on reliable and high performance robotic system for food / biscuit manufacturing line. We propose a design of a vehicle. The robot is capable of picking unbaked biscuits tray and places them into furnace and then after baking it picks the biscuits tray from the furnace. A special gripper is designed to pick and place the biscuits tray with flexibility.

## I. INTRODUCTION

A robot can be defined as a programmable, self-controlled device consisting of electronic, electrical, or mechanical units. More generally, it is a machine that functions in place of a living agent. Robots are especially desirable for certain work functions because, unlike humans, they never get tired; they can work in physical conditions that are uncomfortable or even dangerous; they can operate in airless conditions; they do not get bored by repetition; and they cannot be distracted from the task at hand. This article is based on the research project which is an autonomous robot to be use in food industry. The robot is powerful, reliable and can be use in hot temperature area where a human after working for so long can become sick and exhausted [1]. This research project would also lead to a low cost manufacturing food products as once the robot is implemented it can work repeatedly without any cost. This project will introduce a new era in food industry to use automated machine and robot for more precise, cost effective and reliable work. This robot is able to work in biscuits industry for picking unbaked biscuits tray and placing it in the furnace after backing the biscuits it puts out the tray from furnace and places it on the table. The most apparent reasons that are associated in installing of robotic systems in food industry are;

1) Saving of manpower.
2) Improved quality & efficiency.
3) Ability to work in any hostile environment.
4) Increased consistency & flexibility.
5) Increased yields and reduced wastage.

The robot, which we have been able to make, is completely independent and intelligent. In this robot we will use our own power supply (i.e. a rechargeable battery), which could be mobilized. The sole purpose of micro controller based robot was to propose a design that introduces the idea of automation in food industries. The idea is to reduce manual controlled system, which always needed a human interface. This Robotic system is feasible by small and local industries having small scale production. The system is a reliable, can reduce the cost of production, and reduce the manpower and human workload. A robot can include any of the following components;

a) *Effectors* - like "arms", "legs", "hands", "feet" etc.
b) *Sensors* - parts that act like senses and can detect objects and converts the object`s information into symbols that computer system can understand.
c) *Computer* - the brain that contains instructions to control the robotic system.
d) *Equipment* - this includes tools and mechanical fixtures.
e) *Characteristics-* that makes robot different from regular machinery are that robots usually function by themselves, are sensitive to their environment, adapt to variations in the environment or to errors in prior performance, are task oriented and often have the ability to try different methods to accomplish a task [2].

Common industrial robots are generally heavy rigid devices limited to manufacturing. They operate in precisely structured environments and perform single highly repetitive tasks under preprogrammed control [14].

## II. RESEARCH OBJECTIVES

The main objective for this study was;
a) To increase the manufacturing capacity for local food industries.
b) To increase the labor productivity by the redistribution of laborers in the industries.
c) Reducing the cost factor and manufacturing time of a product.
d) To eliminate the manual based tasks and operations.

## III. INSTALLATIONS AND TECHNIQUES

*A. Microcontroller:*

A micro controller is an inexpensive single chip computer. Single chip computer means that the entire computer system lies within the confines of the integrated circuit chip. The micro controller on the encapsulated sliver of silicon has features similar to those of our standard personal computer [10]. Primarily, the micro controller is capable of storing and running a program. The micro controller contains a CPU, RAM, ROM, I/O lines, serial and parallel ports, timers and sometimes other built in peripherals as A/D (analog to digital) and D/A (digital to analog) converters.

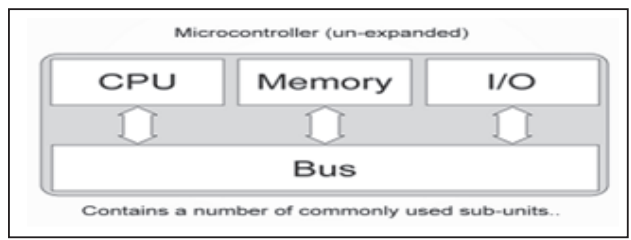

Figure 3.0: Main Components of Microcontroller [9]

Most microcontrollers will also combine other devices such as;
- A Timer module to allow the micro controller to perform tasks for certain time periods [7].
- A serial I/O port to allow data to flow between the micro controller and other devices such as a PC or another micro controller [7].
- An ADC to allow the micro controller to accept analogue input data for processing [7].

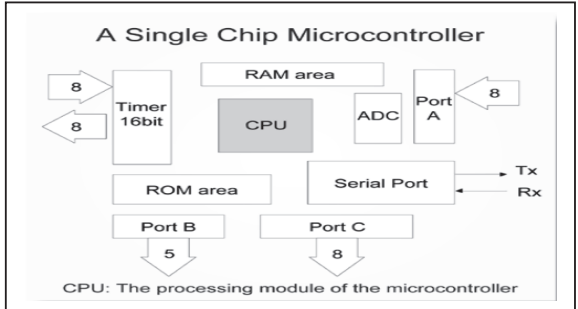

*Figure 3.1: Single Chip Microcontroller [9]*

The above figure illustrates a typical micro controller device and the different sub units integrated onto the micro controller microchip. The heart of the micro controller is the CPU core.

*B. DC Motor:*

Direct current motor is designed to run on DC electric power. The pure DC designs are Michael Faraday's homo-polar motor (which is uncommon), and the ball bearing motor, which is so far a novelty [15]. The most common DC motor types are;

1. Brushed DC motors    2. Brushless DC motors

*1) Brushed DC motors:* The classic DC motor design generates an oscillating current in a wound rotor with a split ring commutator, and either a wound or permanent magnet stator. A rotor consists of a coil wound around a rotor which is then powered by any type of battery [15]. Many of the limitations of the classic commutator DC motor are due to the need for brushes to press against the commutator. This creates friction. At higher speeds, brushes have increasing difficulty in maintaining contact. Brushes may bounce off the irregularities in the commutator surface, creating sparks. This limits the maximum speed of the machine. The current density per unit area of the brushes limits the output of the motor. The imperfect electric contact also causes electrical noise. Brushes eventually wear out and require replacement, and the commutator itself is subject to wear and maintenance [15].

*2) Brushless DC motors:* Some of the problems of the brushed DC motor are eliminated in the brushless design. In this motor, the mechanical "rotating switch" or commutator / brush gear assembly is replaced by an external electronic switch synchronized to the rotor's position [13]. Brushless motors are typically 85-90% efficient, whereas DC motors with brush gear are typically 75-80% efficient [13]. Midway between ordinary DC motors and stepper motors lays the realm of the brushless DC motor. Built in a fashion very similar to stepper motors, these often use a permanent magnet external rotor, three phases of driving coils, one or more Hall Effect sensors to sense the position of the rotor, and the associated drive electronics. The coils are activated, one phase after the other, by the drive electronics as cued by the signals from the Hall effect sensors [13]. In effect, they act as three-phase synchronous motors containing their own variable-frequency drive electronics. A specialized class of brushless DC motor controllers utilizes EMF feedback through the main phase connections instead of Hall Effect sensors to determine position and velocity [13]. These motors are used extensively in electric radio-controlled vehicles. When configured with the magnets on the outside, these are referred to by mode lists as outrunner motors [13].

## IV. PRINCIPLE OF OPERATION

In any electric motor, operation is based on simple electromagnetism. A current-carrying conductor generates a magnetic field; when this is then placed in an external magnetic field, it will experience a force proportional to the current in the conductor, and to the strength of the external magnetic field. As you are well aware that opposite polarities attract, while like polarities repel [9]. The internal configuration of a DC motor is designed to harness the magnetic interaction between a current-carrying conductor and an external magnetic field to generate rotational motion [9].

*1) Stepper Motor:* A stepper motor is a brushless, synchronous electric motor that can divide a full rotation into a large number of steps [11]. The motor's position can be controlled precisely, without any feedback mechanism. Stepper motors are similar to switched reluctance motors, which are very large stepping motors with a reduced pole count, and generally are closed-loop commutated [11].

*2) Fundamental Operation:* Stepper motors operate differently from normal DC motors, which rotate when voltage is applied to their terminals [8]. Stepper motors, on the other hand, effectively have multiple "toothed" electromagnets arranged around a central gear-shaped piece of iron [8]. The electromagnets are energized by an external control circuit, such as a microcontroller. To make the motor shaft turn, first one electromagnet is given power, which makes the gear's teeth magnetically attracted to the electromagnet's teeth [8]. When the gear's teeth are thus aligned to the first electromagnet, they are slightly offset from the next electromagnet [8]. So when the next electromagnet is turned on and the first is turned off, the gear rotates slightly to align with the next one, and from there the process is being repeated. Each of those slight rotations is called a "step" [8]. In that way, the motor can be turned to a precise angle.

*3) Stepper motor characteristics:* Stepper motors are constant-power devices *(Power = Angular Velocity X Torque)* [8]. The torque curve may be extended by using current limiting drivers and increasing the driving voltage [8]. Steppers exhibit more vibration than other motor types, as the discrete step tends to snap the rotor from one position to another. This vibration can become very bad at some speeds and can cause the motor to lose torque [8]. The effect can be mitigated by accelerating quickly through the problem speed range, physically damping the system, or using a micro-stepping driver. Motors with a greater number of phases also exhibit smoother operation than those with fewer phases [8]. There are two main types of stepper motors;

a) Permanent Magnet Stepper    b) Hybrid Stepper

a) *Permanent Magnet Stepper:* A permanent magnet stepper motor has a cylindrical permanent magnet rotor. The stator usually has two windings [6]. The windings could be center tapped to allow for a uni-polar driver circuit where the polarity of the magnetic field is changed by switching a voltage from one end to the other of the winding [6]. A bi-polar drive of alternating polarity is required to power windings without the center tap. A pure permanent magnet stepper usually has a large step angle. Rotation of the shaft of a de-energized motor exhibits detent torque [7]. If the detent angle is large, say 7.5º to 90º, it is likely a permanent magnet stepper rather than a hybrid stepper [6].

b) *Hybrid Stepper:* The hybrid stepper motor combines features of both the variable reluctance stepper and the permanent magnet stepper to produce a smaller step angle [4]. The rotor is a cylindrical permanent magnet, magnetized along the axis with radial soft iron teeth (Figure: 4.0). The stator coils are wound on alternating poles with corresponding teeth [4]. There are typically two winding phases distributed between pole pairs. This winding may be center tapped for uni-polar drive. The center tap is achieved by a bifilar winding, a pair of wires wound physically in parallel, but wired in series [4]. The north-south poles of a phase swap polarity when the phase drive current is reversed [4]. Bipolar drive is required for un-tapped windings.

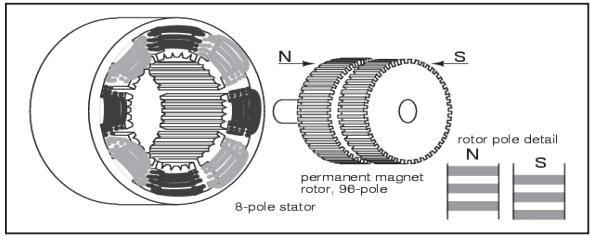

*Figure 4.0: A hybrid Stepper Motor [4]*

## V. MODEL AND DESIGN

*A. Layout of the Robot*:

The mobile pick and place robotic sysem as shown in figure 4.1: can be represented by three basic subsystems as;
1. Moveable base   2. Rotational manipulator   3. Magnetic gripper unit.

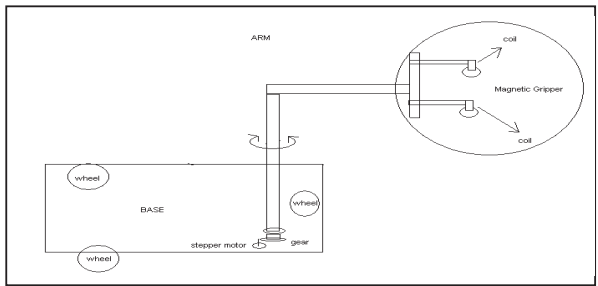

*Figure 5.0: Layout of the Robot*

*1. Movable Base:* The robot base is three wheel vehicles. The two backward wheels are derived by DC motor through mechanical gears and controlled by controller through H-bridge drive circuit [3]. It can move in forward and backward direction. The front wheel is fixed to support the base [3].

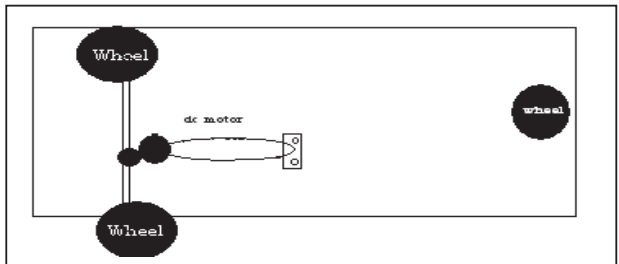

*Figure 5.1: Base of the Robot*

*2. Rotational Manipulator:* The rotational robotic arm consists of a stepper motor that moves in step angles to $360^0$ by giving logic through controller [5]. A stepper motor drive circuit drives the motor. The manipulator is one joint and has one degree of freedom [5].

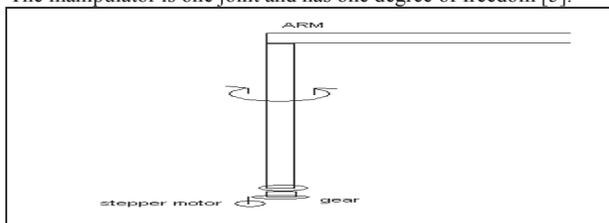

*Figure 5.2: The Manipulator [7]*

*3. Magnetic Gripper Unit:* We have design a special type of gripper that is based on electromagnetic effect. The gripper consists of two-inductor coil on each side, which operates on 12 V DC signals [5]. When 12V DC signal is applied, from the battery, the coil in the gripper energizes and induces an electromagnetic field around it [12]. The induce electromagnetic field around the gripper picks up the metallic biscuit tray from the desired location and places it on the target place, with flexibility, when the 12V DC signal is removed [13].

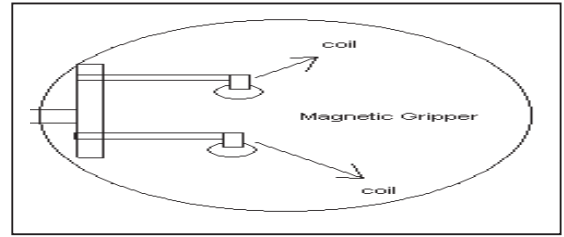

*Figure 5.3: Magnetic Gripper [5]*

*B. Special Application:*

The gripper can be suitable in areas where we have to pick and place metal object or toll in many other industries [12].

*C. Operating the Robot:*

First the manipulator arm moves $90^0$ and picks the biscuit tray through the magnetic gripper from the table. Then the robot moves forward to the furnace and places the biscuit tray into the furnace. It waits for sometime till the biscuits baked and again picks the biscuit tray from the furnace and places it on the table.

## VI. CIRCUIT IN THE ROBOTIC SYSTEM

- *DC Motor Drive Circuit:* The H-bridge as shown in figure 6.0, is use to drive the DC-motor forward and backward.

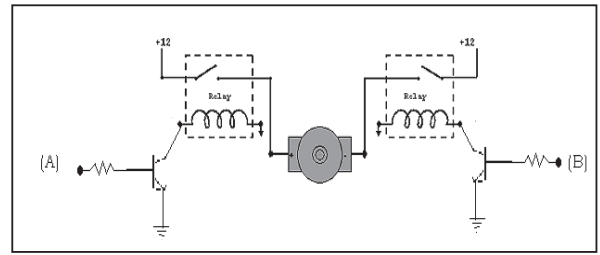

*Figure 6.0: H-bridge circuit [2]*

The +12V DC power is applied from the battery to the H-bridge circuit, and controlled through the controller by applying logics.

| A | B | FUNCTION |
|---|---|---|
| 1 | 0 | Forward |
| 0 | 1 | Reverse |
| 1 | 1 | Stop |
| 0 | 0 | Stop |

*Table 6.1: H-bridge circuit Boolean Logic*

In table 6.1, when logic 1, 0 is applied to terminal A and B, the motor moves in forward direction and when logic 0, 1 is applied it moves in reverse direction respectively. The motor will be in rest state when logic 0, 0 or 1, 1 is applied.

- **Stepper Motor Drive Circuit:** The circuit in figure 6.2, is use to drive stepper motor. The stepper motor has 6 wire in which two are use for VCC and GND [3].

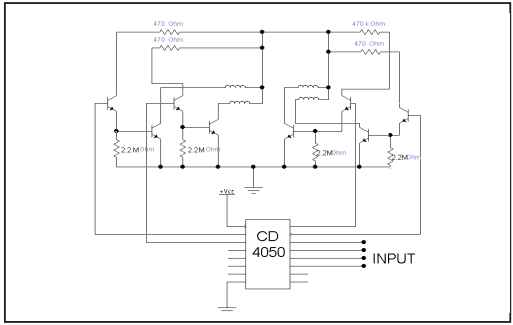

*Figure 6.2: Stepper motor drive circuit [1]*

The remaining 4 wire is used to provide logic through the below shown circuit from the controller. The logic given to the drive circuit to move motor in step angles via controller is given in table 6.3;

| X | $\overline{X}$ | Y | $\overline{Y}$ | Step Angle |
|---|---|---|---|---|
| 0 | 1 | 0 | 1 | 0° |
| 1 | 0 | 0 | 1 | 90° |
| 1 | 0 | 1 | 0 | 180° |
| 0 | 1 | 1 | 0 | 270° |

*Table 6.3: Stepper motor drive circuit Boolean logic*

- **Inputs:** These ICs are unusual because their gate inputs can withstand up to +15V even if the power supply is a lower voltage.
- **Outputs:** These ICs are unusual because they are capable of driving 74LS gate inputs directly. To do this they must have a +5V supply (74LS supply voltage). The gate output is sufficient to drive four 74LS inputs. *NC = No Connection (a pin that is not used).*

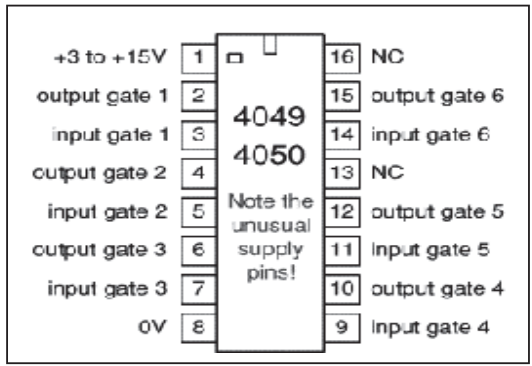

*Figure 6.4: Pin configuration of 4049 / 4050 buffers [3]*

## VI. RESULTS

*1. Degree of Freedom:* The point at which the robot manipulator rotates is called joint or axis [12]. The number of joints in an industrial robot determines its Degree of freedom of motion. Our robot has one movement of axis so it posses" *one degree of movement*" [5], which means it has only one movement that is rotational.

*2. Work space:* The reach of the robot is defined as the workspace or work envelop of the system. All programmed points within the reach of robot are part of the workspace. The workspace shape of the project robot is cylindrical as it can rotate in $360^0$ angles [4].

*3. Robot Motion:* The robot has two basic movements;

   1. The base travel movement.   2. The arm rotational movement.

*4. End- Effector:* The end-Effector is the hand connected to the robot arm. It is different from the human hand. The end- Effector gives the robotic system the flexibility necessary for the operation of the robot. The end- Effector of the project is a magnetic gripper and is driven by 12V DC source [4].

*5. Pay Load:* Payload is the load capacity of the robot. The project robot has a low payload of 200g.

*6. Accuracy:* The accuracy of a robot describes how closely a robot can position its manipulator. The accuracy rate of the project is 70% due to mechanical gears [7].

*7. Actuations:* The method of driving the robot axes is called actuation. The actuation technique used in the robot is electrical [7].

*8. Dimensions of Robot:* Length of Robot is: 21 inches; Height of Robot is: 17 inches; Width of Robot is: 10 inches.

## VII. FUTURE IMPLEMENTATIONS

We have done our best efforts to make the project feasible, simple and reliable for the local industrial usage. There can be modification in this robotic system that can be more efficient and effective like;

- The robotic system can be modified by implementing vision system and artificial intelligence to avoid obstacles in between the path.
- The vision system can also be implemented to check whether the biscuits are properly baked or not.
- The microcontroller programming can be replaced by PLC, as a new technology.
- Making magnetic gripper more powerful can increase the payload property of the robot.
- The gripper can be modified for different operations in different industries.
- Advancement can be made in the robot by implanting temperature sensors.

## APPENDIX - A

*Microcontroller Interface Circuit:*

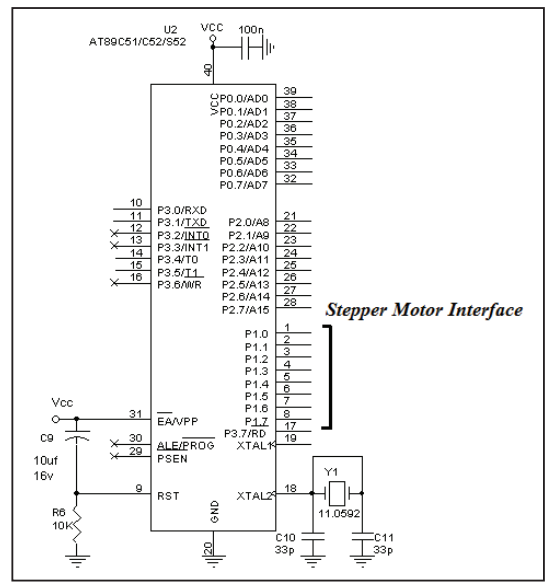

ACKNOWLEDGMENT

We thank the anonymous reviewers for invaluable guidance on improving our presentation of this material. We are also thankful to our beloved parents for their love, trust and support.

**Authors:**

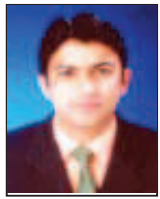

**Mir Sajjad Hussain Talpur,** has received Master`s in Information Technology, & another Master`s degree in (Science & Technology Policy) from Mehran University of Engineering and Technology (MUET) in 2009 & Currently PhD Scholar in School of Information Science & Engineering, Central South University Changsha, China, in 2010. He is Lecturer in IT Department Sindh Agriculture University, Tandojam Hyderabad, Pakistan, since 2004, having additional assignment of Faculty Coordinator, program coordinator of Quality Enhancement Cell. His research interests include Internet of Things, Security & Privacy Issues, Software Engineering, Computer networking & Intelligent Systems. His research is also supported by talents programs including Program for New Century Excellent Talents in University.

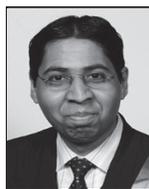

**Murtaza Hussain Shaikh,** has earned Master`s student in (Information Systems Engineering) from Norwegian University of Science and Technology (NTNU), Norway in 2011 and another Master`s degree in (Science and Technology Policy) from Mehran University of Engineering and Technology (MUET), Pakistan in 2009. He got his Bachelor`s degree in (Software Engineering) from University of Sindh, Pakistan in 2006 and another in (Arts &Economics) from Allama Iqbal Open University, Islamabad, Pakistan in 2008. His current research interests include IS Security, Technological Law and Policy Making of Computer Organization, IT Systems operations and Maintenance, Cyber Crimes and Cyber Security issues. He has a research and teaching experiences in Computer Science in different Pakistan`s Universities and Colleges. He has authored and co-author many research articles in Local and International Journals and Conferences. He is also a member of International / local professional associations and committees like PEC, PCB, IACSIT, Singapore and IJPA – USA.